\documentclass[10pt,twocolumn]{IEEEtran}
\usepackage{bbding}
\usepackage{graphicx}
\usepackage{amsmath}
\usepackage{latexsym}
\usepackage[table]{xcolor}
\usepackage{booktabs}
\usepackage{caption}
\usepackage{subcaption}
\usepackage{multirow}
\usepackage{amsthm}
\usepackage{algorithm}
\usepackage[noend]{algpseudocode}

\theoremstyle{definition}

\newcommand*\rot{\rotatebox{90}}

\newlength\myindent
\makeatletter
\def\BState{\State\hskip-\ALG@thistlm}
\makeatother

\ifCLASSINFOpdf
\else
\fi
\begin{document}
\title{Deep Learning for Secure Mobile Edge Computing}
\author{
\IEEEauthorblockN{Yuanfang Chen\IEEEauthorrefmark{1}, Yan Zhang\IEEEauthorrefmark{2}\Envelope, Sabita Maharjan\IEEEauthorrefmark{2}}\\
\IEEEauthorblockA{
\IEEEauthorrefmark{1}Cyberspace School, Hangzhou Dianzi University, China\\
\IEEEauthorrefmark{2}University of Oslo, Norway}
\thanks{Corresponding author: Yan Zhang}
}
\maketitle

%======================================================================================================================================================================================================================================

\begin{abstract}
Mobile edge computing (MEC) is a promising approach for enabling cloud-computing capabilities at the edge of cellular networks.  Nonetheless, security is becoming an increasingly important issue in MEC-based applications.  In this paper, we propose a deep-learning-based model to detect security threats.  The model uses unsupervised learning to automate the detection process, and uses location information as an important feature to improve the performance of detection.  Our proposed model can be used to detect malicious applications at the edge of a cellular network, which is a serious security threat.  Extensive experiments are carried out with 10 different datasets, the results of which illustrate that our deep-learning-based model achieves an average gain of 6\% accuracy compared with state-of-the-art machine learning algorithms.
\end{abstract}

\IEEEpeerreviewmaketitle

%======================================================================================================================================================================================================================================

\section{Introduction}
\label{sec:introduction}
Mobile edge computing (MEC) is often referred to as a plausible approach to reduce end-to-end delay by moving cloud-computing capabilities to the edge~\cite{liang2017mobile}.  In addition, MEC can provide location awareness services for cellular-network-based applications.  While MEC is becoming an increasingly popular computing paradigm for dynamic systems/scenarios, new types of vulnerabilities, especially those that can come from the edge network itself, are also becoming serious concerns.  For instance, when an edge customer connects with the closest edge computing device to access a service in a certain location, an attacker can tamper with the address of the currently connected edge computing device stored with the edge customer, to possibly interrupt the connection.  Such manipulations are location-aware.  Detecting such attacks and associating them with the locations of their origins are both important.

\begin{figure*}[!ht]
  \centering
  \includegraphics[width=4.5in]{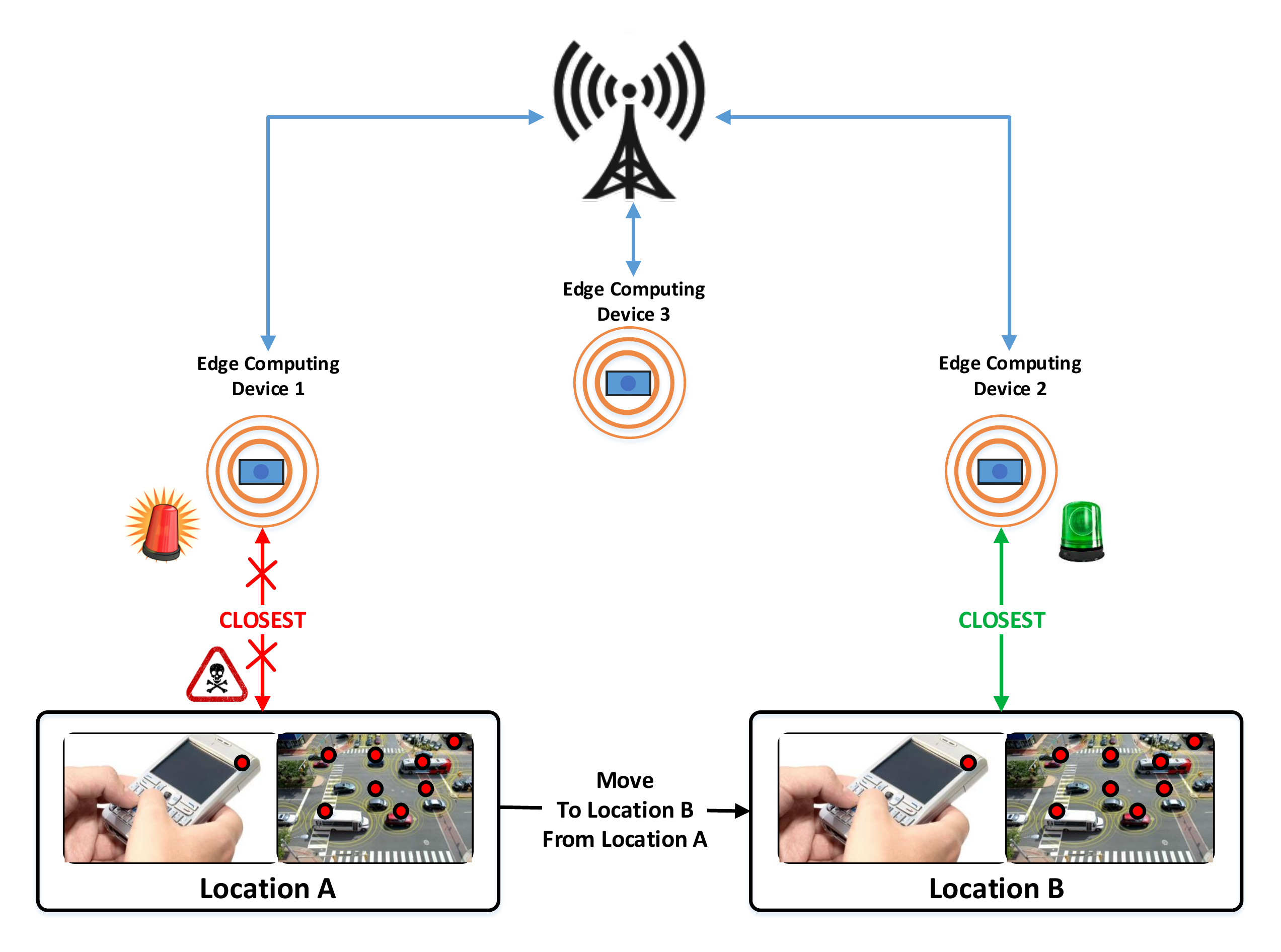}\\
  \caption{In mobile edge computing, an edge customer is connected to the closest edge computing device to obtain a service.  As the edge customer moves, it will connect to a different edge computing device.  Location information is a feature to be incorporated in the detection algorithm to enhance the performance of detecting attacks in the location-aware MEC environment.}
  \label{fig:problem}
\end{figure*}

Moreover, such attacks can result in extra energy consumption.  Tampering with the address of the closest connected edge computing device to disconnect the edge customer from its closest edge computing device is a typical example of such an attack.  The edge customer in this case has to connect to a remote edge computing device, thus resulting in extra energy consumption of the network connection.  Based on the above description, the research problem is illustrated in Figure~\ref{fig:problem}.  An edge customer in location A attempts to connect to the closest edge computing device 1, but the connection is ineffective because of the attack.  As the customer moves, it attempts to connect to the closest edge computing device 2 in location B.  Because the attack takes place in location A, not in location B, the connection to edge computing device 2 is successful.  To detect this attack, the detection method needs location information to automate the detection process.

In order to detect such attacks, we introduce a deep-learning-based model.  As an example, the introduced model is used to detect the malicious application on Android devices~\cite{hou2017hindroid}, which is a kind of serious security threat for mobile devices in the MEC environment.  Using machine learning algorithms to detect such malicious attacks, requires extracting features from the applications.  The state of the art provides quite a good volume of work in this direction.  DynaLog~\cite{alzaylaee2016dynalog} is a dynamic analysis framework that automatically extracts dynamic behaviors (features) of applications for further processing.  The authors in~\cite{shabtai2012andromaly} presented a dynamic framework called Andromaly that applies several different machine learning algorithms to classify applications.  MADAM~\cite{dini2012madam} is also a dynamic analysis framework that uses machine learning to classify applications.  It extracted 13 features at the user and kernel levels.  Marvin~\cite{lindorfer2015marvin} combined static and dynamic analysis and applied machine learning techniques to assess the risk of unknown applications.  Crowdroid~\cite{burguera2011crowdroid} is a cloud-based machine learning framework for malicious application detection.

However, the existing work has not considered the location information related to the attack that can be exploited to learn more about it, and thus detect the attack in an autonomous manner.  In this paper, we propose a deep-learning-based model to detect malicious applications while also incorporating the location information into the detection framework.  For adapting the dynamic computing paradigm, this model uses a deep belief network to automatically extract and learn the features of malicious applications, and then, by classification technique, the model can detect whether an application is malicious or not in a certain location with the help of the features.  We compare the performance of our proposed model with four widely adopted machine learning algorithms, and illustrate the gain offered by our model.

The remainder of this paper is organized as follows.  In Section~\ref{sec:machine_learning_based_malware_detection}, we review the existing efforts on machine-learning-based malicious application detection. In Section~\ref{sec:framework}, we describe the proposed deep-learning-based model.  In Section~\ref{sec:performance_comparison}, we compare the performance of four widely used machine learning algorithms with the proposed model.  Section~\ref{sec:conclusion} concludes the paper.

%======================================================================================================================================================================================================================================

\section{Machine-Learning-Based Detection: State of the Art}
\label{sec:machine_learning_based_malware_detection}
Existing malicious application detection schemes can mainly be classified into two categories.
\begin{enumerate}
  \item Static analysis examines the source code of applications without executing them.  It analyzes the static features extracted from the source code to learn malicious patterns of applications.  With such learning, it is possible to know the malicious behaviors, and then to avoid malicious attacks.  The static features of applications include permissions, used hardware components, and data-flow usage.
  \item Dynamic analysis examines applications in a running environment, and monitors the dynamic behavior of applications.  The dynamic features include network connections, system calls, and resource usage.  Such analysis is conducted in the running status of an application to analyze the dynamic behavior of the application, for instance, analyzing which networks an application will connect during the period of running.
\end{enumerate}

In both kinds of schemes, the data are collected to train machine learning classifiers to model benign and malicious features of the applications, and then the trained model detects malicious applications.

\subsection{Static Analysis}
DERBIN~\cite{arp2014drebin} is a method for the detection of Android malicious applications on smart phones.  It is a static analysis framework that extracts a set of features from the document AndroidManifest.xml and the disassembled code of applications.  In DERBIN, a support vector machine was applied to learn the difference between benign and malicious applications.  Droidmat~\cite{wu2012droidmat} detects malicious applications by analyzing AndroidManifest.xml and tracing system calls.  It depends on static analysis for the Android applications' permissions, components, intent messages, and API calls.  Droidmat extracts different features from the document AndroidManifest.xml, and then it applies the k-means algorithm to model malicious applications, while the number of clusters is determined by singular value decomposition.  In~\cite{sanz2012automatic}, the authors deployed decision-tree-, k-nearest-neighbor-, Bayesian-network-, random-forest-, and support-vector-machine-based algorithms for automatic Android application categorization and malicious application detection.  With that, the authors indicated that the Bayesian network is a better classifier compared with those based on a random forest and decision tree.  In~\cite{sahs2012machine}, a system was built up by training a support vector machine classifier with the extracted static feature from Android applications, namely permissions.

\subsection{Dynamic Analysis}
Shabtai~\emph{et al.},~\cite{shabtai2012andromaly} proposed an artificial-neural-network-based system to detect unknown Android malicious applications based on analyzing the applications' permissions and system calls.  In this work, two types of artificial neural networks were used: feedforward neural networks were used for training the model that builds distinguishable patterns between benign and malicious applications, and the training is conducted with the requested permissions of the applications; the recurrent neural networks were used to train the model with the system calls of the benign applications' execution behavior.  Crowdroid~\cite{burguera2011crowdroid} applies k-means on the vectors of system's calls.  It can distinguish the applications that have the same name and version, but different behavior, in the Android platform.  AntiMalDroid~\cite{zhao2011antimaldroid} is an application-behavior-signature-based detection framework that deploys the support vector machine algorithm.  It can detect malicious applications and variants in runtime by analyzing the dynamic behavior of malicious applications, and it can extend the characteristic database of malicious applications dynamically.  Table~\ref{tab:review} shows the overview of existing machine-learning-based schemes for detecting malicious applications.
\begin{table*}[!ht]
\centering
\caption{Machine-learning-based malicious application detection: Summary of state of the art}
\label{tab:review}
%\vspace*{-\baselineskip}
\renewcommand{\arraystretch}{2}
\begin{tabular}{lll}
\bottomrule
\rowcolor{black!15}Approach & Machine Learning Method & Author\\\midrule
\multirow{4}*[2ex]{\rot{\parbox{1.5cm}{\centering Static Analysis}}}
& Support Vector Machine & Arp~\emph{et al.}~\cite{arp2014drebin}\\
& K-Means Algorithm & Dongjie Wu~\emph{et al.}~\cite{wu2012droidmat}\\
& Decision Tree, K-Nearest Neighbours, Bayesian Network, Random Forest, and Support Vector Machine & B. Sanz~\emph{et al.}~\cite{sanz2012automatic}\\
& Support Vector Machine & J. Sahs~\emph{et al.}~\cite{sahs2012machine}\\
\cline{2-3}
\multirow{3}*[2ex]{\rot{\parbox{1.5cm}{\centering Dynamic Analysis}}}
& \cellcolor{black!15}{Two Types of Artificial Neural Networks: Feedforward Neural Networks and Recurrent Neural Networks} & \cellcolor{black!15}{Shabtai~\emph{et al.}~\cite{shabtai2012andromaly}}\\
& K-Means Algorithm & I. Burguera~\emph{et al.}~\cite{burguera2011crowdroid}\\
& Support Vector Machine & M. Zhao~\emph{et al.}~\cite{zhao2011antimaldroid}\\
\bottomrule	
\end{tabular}
\end{table*}

In the existing work, location information is not considered as a feature to conduct detection.  Location information is important for the location-aware computing paradigm, MEC, to improve the detection accuracy.  We propose a deep-learning-based model to detect malicious applications, which can incorporate location information to improve detection performance.  The model uses location information as an important feature that is used, together with other features, as the input of a deep belief network to conduct model training.

We compare our model with other four machine learning algorithms, and these four algorithms have different process features:
\begin{itemize}
  \item Decision tree.  A decision tree incrementally constructs an associated decision tree by breaking down a dataset into smaller subsets.  Even if more data are added and the data size is increased, the construction process is similar.  Such adding and increasing cannot always help increase the accuracy of the decision tree.  In Figure~\ref{fig:ratio-accuracy}, for the decision tree, the accuracy value is smaller at 500/5000 compared with that at 500/4500.
  \item Random forest.  A random forest is constructed by a multitude of decision trees, and outputs the classes (classification) or mean prediction (regression) of the individual trees.  Forest construction can be reduced to the construction of multiple trees, and thus is similar to a decision tree, in that even if more data are added and the data size is increased, it cannot always help to increase the accuracy of the random forest.  For example, for the random forest, the accuracy value is smaller at 500/2500 compared with that at 500/2000.
  \item Softmax regression.  Softmax regression parameters are trained to minimize a cost function, and it is a single-layer artificial neural network for linearly separable datasets; however, the datasets used in our study are not linearly separable.
  \item Support vector machine.  In a support vector machine, the kernel function is the core of determining the performance, not data-based training.
\end{itemize}

The deep-learning-based model mainly offers two advantages: (i) Improvement of detection accuracy, and (ii) flexibility.  Regarding the improvement of detection accuracy, the proposed deep-learning-based model benefits from data-based model training.  The data from the application have enough information to reflect the complicated relationship among all parameters of the application.  Regarding flexibility, in the deep-learning-based model it is easy to add new features to satisfy the requirements from more complex application scenarios, e.g., adding the location feature as an input parameter.  It is easy to add new modules to develop new functions, e.g., using the unsupervised learning function of a deep belief network to automate the feature learning process.  This means that it is not necessary to manually set the relationship between feature values and a malicious application, e.g., if feature A=1, the application is malicious; otherwise, it is benign.

%======================================================================================================================================================================================================================================

\section{Deep-Learning-Based Model}
\label{sec:framework}
We design a deep-learning-based model to detect security threats, and as an example, we use our model to detect the malicious Android application which is a kind of serious security threat in cellular networks.  For instance, an Android application includes malicious code to threaten network security.

This model consists of two components: a feature preprocessing engine and a malicious application detection engine.

\textbf{Feature Preprocessing Engine.}  To characterize the applications (both malicious and benign), static and dynamic analyses are conducted to preprocess the features for each application.  The features fall into one of three types: required permissions, sensitive APIs, and dynamic behavior.  Required permissions and sensitive APIs are processed through static analysis, and the dynamic behavior is processed through dynamic analysis.

In the feature preprocessing engine, APK files from the edge devices of MEC are unpacked first, and then the feature elements that will be used as the input of the malicious application detection engine are extracted from the unpacked files.  Taking the \emph{permissions} feature as an example, for a special application, if a kind of ``permission'' is used, it is recorded as 1; otherwise, it is recorded as 0.  This creates a two-dimensional array of bits.  Lastly, these two-dimensional arrays are stored into JSON files.

\textbf{Malicious Application Detection Engine.}  There are two parts in this engine: deep-belief-network-based feature learning and softmax-function-based prediction output.

\begin{figure*}[!ht]
  \centering
  \includegraphics[width=5in]{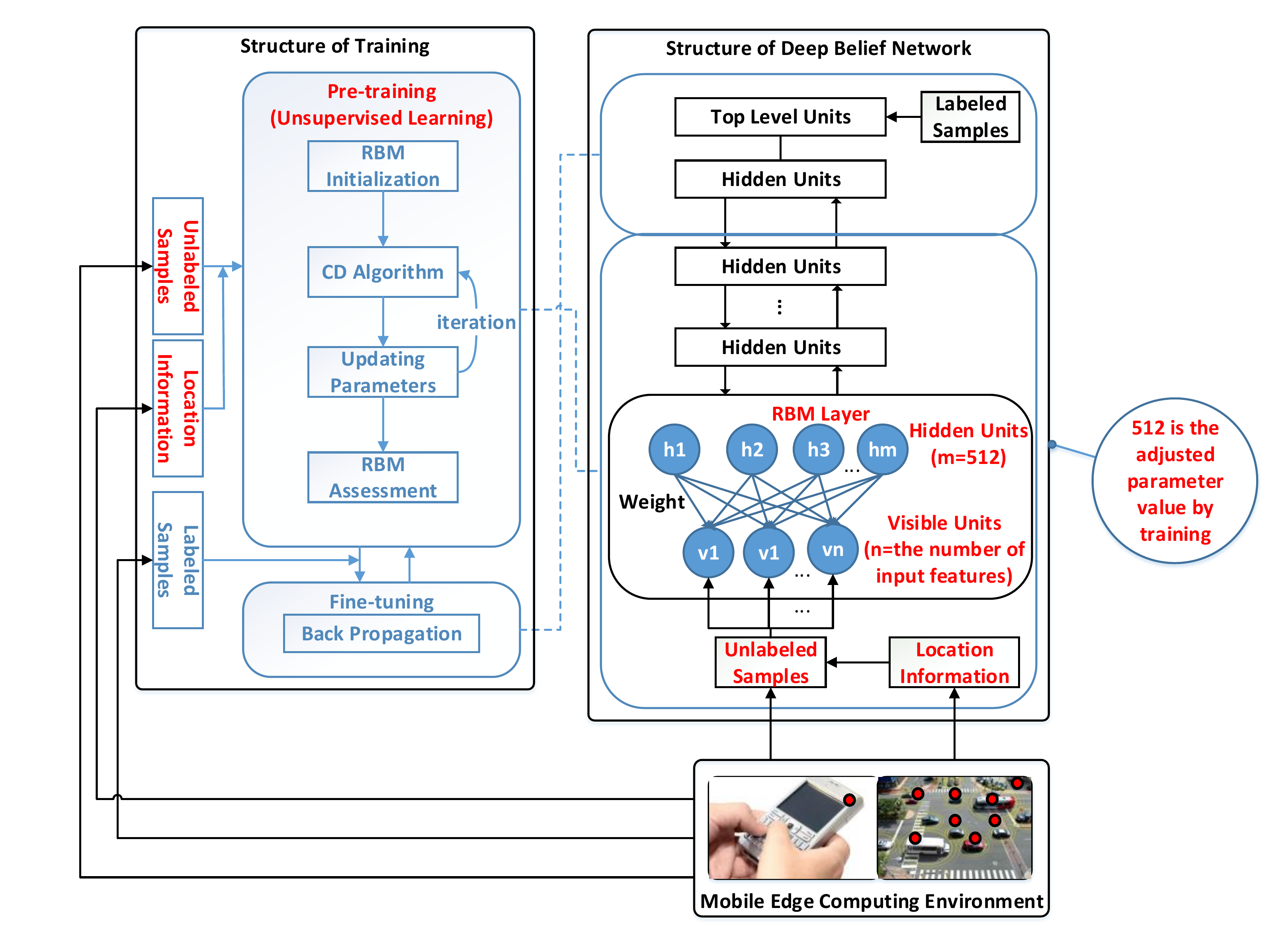}
  \caption{Structure of the deep belief network, and network training with the samples from the MEC environment.}
  \label{fig:structure_training_DBN}
\end{figure*}

This engine develops a deep belief network to automatically learn the features of malicious applications.  The structure of the deep belief network is illustrated in Figure~\ref{fig:structure_training_DBN} (right half).

The deep belief network can be viewed as a composite of simple and unsupervised networks such as restricted Boltzmann machines (RBMs)~\cite{hinton2009deep}~\footnote{An RBM is a generative stochastic artificial neural network that can learn probability distribution over its set of inputs, and it is comprised of one-layer visible units, one-layer hidden units, and weights between the visible units and the hidden units.}. Multi-RBMs are stacked together, and the output from the hidden layer of the prior RBM is provided as the input of the next RBM.  Moreover, the input of the first RBM layer is unlabeled samples from the MEC environment\footnote{The label that is used to label a feature denotes that the feature is from a malicious or benign application.}.  A set of features is used as the unlabeled samples to be fed as the input to the visible units of the RBM layer.  The features are extracted from the document AndroidManifest.xml and the log files of edge devices, which are used to analyze the difference of the behavior between malicious and benign applications.  In addition to location information, the following features are used:
\begin{flushleft}
INVOKE\_INTERNAL\_HANDLER, ACCESS\_NORTON\_SECURITY, READ\_FRAME\_BUFFER, WRITE\_GMAIL, READ\_PROFILE, WAVE\_LOCK, ACCESS\_WIMAX\_STATE, VIBRATION, ACCESS\_GPS, ACCESS\_COARSE\_UPDATES, READ\_PROFILE, ACCESS\_WIMAX\_STATE, ACCESS\_COARSE\_UPDATES, INSTALL\_THEME, RECEIVE\_BROADCASTS, ACCESS\_GPS, READ\_INPUT\_STATE, WAVE\_LOCK, ACCESS\_NORTON\_SECURITY, and UNLOCK.
\end{flushleft}
On the top of the deep belief network, there are hidden and top-level units that construct the back-propagation network, which can be used to achieve the fine-tuning of deep learning, and labeled samples are used as input to conduct supervised learning.

There are two phases in the training of this deep belief network:
\begin{itemize}
  \item [i)] Unsupervised pre-training with unlabeled samples (location information has been added into these samples).  The contrastive divergence (CD) algorithm~\cite{carreira2005contrastive} is applied to develop the RBM based training, and its main steps are summarized in Algorithm~\ref{alg:cd}.
  \begin{algorithm}[!ht]
    \caption{Contrastive Divergence}
    \label{alg:cd}
    \begin{algorithmic}[1]
    \Procedure{:}{}
    \State initialize visible units;
    \State \parbox[t]{\dimexpr\linewidth-\algorithmicindent}{update hidden units according to the input from the visible units;\strut}
    \State \parbox[t]{\dimexpr\linewidth-\algorithmicindent}{reconstruct the visible units following the output of the hidden units;\strut}
    \State \parbox[t]{\dimexpr\linewidth-\algorithmicindent}{re-update the hidden units according to the reconstructed visible units;\strut}
    \State \parbox[t]{\dimexpr\linewidth-\algorithmicindent}{update the weights between visible units and hidden units;\strut}
    \EndProcedure
    \end{algorithmic}
  \end{algorithm}
  \item [ii)] Supervised fine-tuning with labeled samples.  Back propagation is used to finely tune the pre-trained parameters that were trained in the pre-training phase.  Back propagation is a method used in artificial neural networks to calculate the error contribution of each neuron.  It is used by an optimization model to tune the weight of each neuron.  It calculates the gradient of the loss function~\footnote{For back propagation, the loss function calculates the difference between the network output and its expected output after a case propagates through the network.}, and it is commonly used in the gradient descent optimization model.  It is also called backward propagation of errors, because the error is calculated at the output and distributed back through the network layers.  Back propagation requires a known output for each input value, so it is considered as a supervised learning method.  The input of the back propagation is the labeled samples from the MEC environment.
\end{itemize}

The entire deep belief network is trained when learning for the final hidden layer is executed.  The structure of training the deep belief network is shown in the left-hand part of Figure~\ref{fig:structure_training_DBN}.

With data-based model training, the model will be more suitable for the current application where the data come from.  In a real-world application, there is a large number of interdependent parameters, making it either impossible or at least extremely difficult to model a system with simple formulations while also capturing the essence of the interdependencies.  On the other hand, the data from the application have enough information to reflect the complicated relationship among all interdependent parameters, thereby using the data facilitates training of an artificial neural network to form a rich model that can properly reflect the dependencies.

The strengths of the model are listed as follows:
\begin{itemize}
  \item [i)]  We developed a deep learning model to capture the non-linear and complicated relations between features and malicious applications.  Compared with the other four algorithms, deep learning shows its most powerful intrinsic property, named feature learning~\cite{lecun2015deep}.  The process of learning the relationship between features and a malicious application is automated with unsupervised learning.  However, in the four state-of-the-art algorithms, designing a good feature adapter requires a considerable amount of effort.  To this end, automated learning of the features using a general purpose learning algorithm, is the key advantage of deep learning.
  \item [ii)]  The deep-learning architecture has a multi-layer stack of modules to compute the non-linear input-output mapping.  Each module in the stack transforms its input to increase both the selectivity and invariance of the representation.  With multiple non-linear layers, it is possible to develop an extremely intricate function of input.
\end{itemize}

In the output layer of the malicious application detection engine, the output of the deep belief network is used as the input of the softmax function to obtain the output.  The output of the softmax function can be used to represent a categorical distribution.  It is a probability distribution over different possible outcomes.  In our case, the output represents a category: output=1 means malicious and output=0 means benign.

In the datasets that are used for experiments of this paper, we have 500 samples from 500 malicious applications, for instance, \emph{air.BuyukCevsen}, \emph{air.CevsenulKebir}, and \emph{air.com.intersanyazilim.RMZAndroid}. Our model can be used in these applications at least to conduct detection.

%======================================================================================================================================================================================================================================

\section{Evaluation and Comparison}
\label{sec:performance_comparison}
\subsection{Experimental Setup}
We used 10 different datasets for evaluating the performance of our model.  The datasets used include 500 malicious application samples, and the number of benign applications is varied as 500, 1000, 1500, 2000, 2500, 3000, 3500, 4000, 4500, and 5000.  For each dataset, the ratio of the data samples used to train and test are 0.8 and 0.2, respectively.  Moreover, we compared the performance of our proposed deep-learning-based model with four standard machine learning algorithms widely used for detecting malicious applications: decision tree, random forest, softmax regression, and support vector machine.  Each algorithm was run 10 times to obtain average values for the performance.

For the proposed deep-learning-based model, we deployed an RBM with 512 hidden units (in its hidden layer; the value 512 is obtained by training based adjustment), and the number of visible units in the RBM is equal to the number of input features.  Moreover, to train the model, the number of iterative epochs of the pre-training phase was 200, and the number of iterative epochs of the fine-tuning phase was set as 200.

\subsection{Experimental Results}
\begin{figure}[!ht]
  \centering
  \includegraphics[width=3.5in]{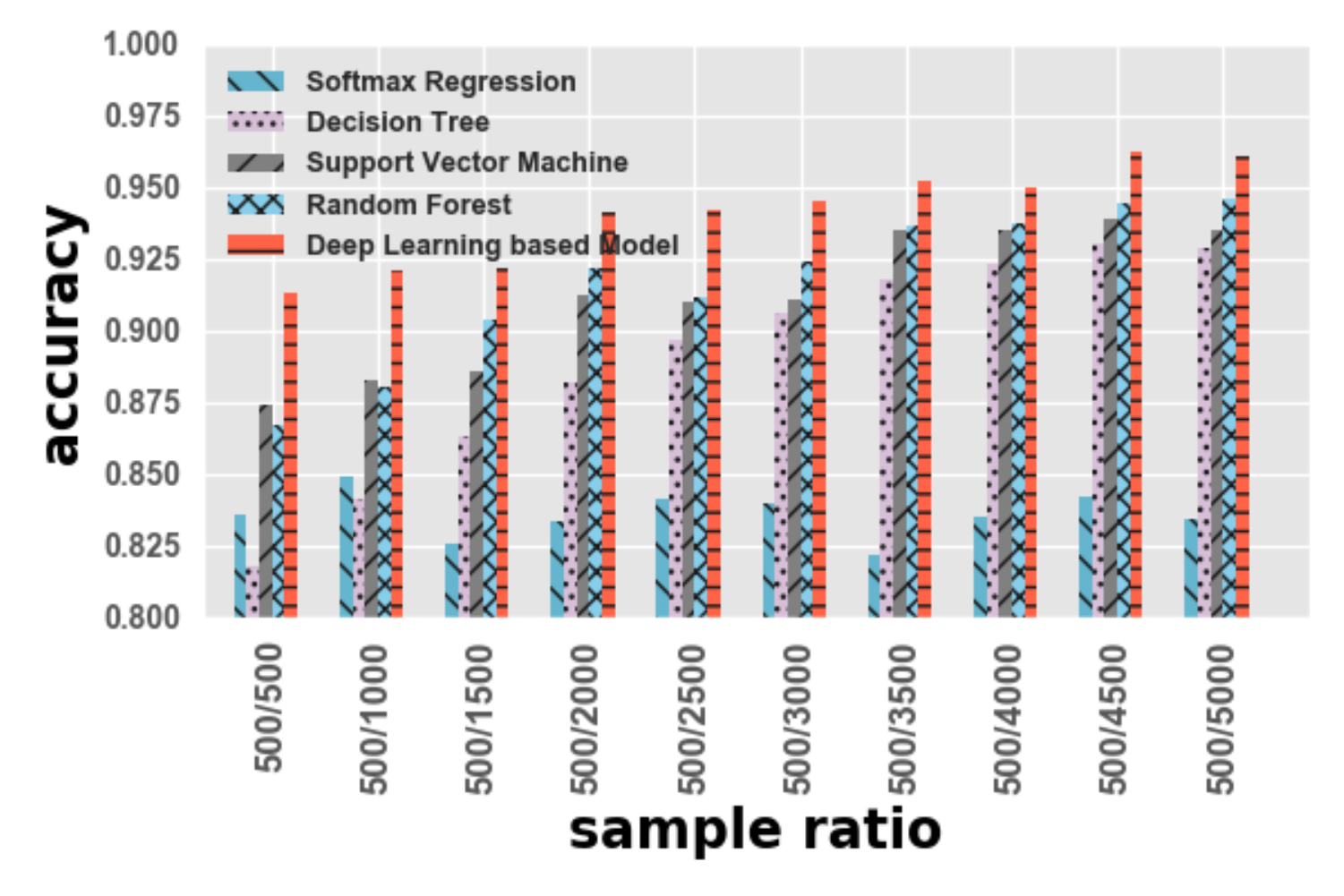}
  \caption{Accuracy comparison of five algorithms on 10 datasets.  The result on a dataset is the average of 10 runs for each algorithm.  The sample ratio is the proportional relation between malicious and benign samples.  For example, 500/1000 means that there are 500 malicious samples and 1000 benign samples.}
  \label{fig:ratio-accuracy}
\end{figure}

\begin{figure}[!ht]
  \centering
  \includegraphics[width=3.5in]{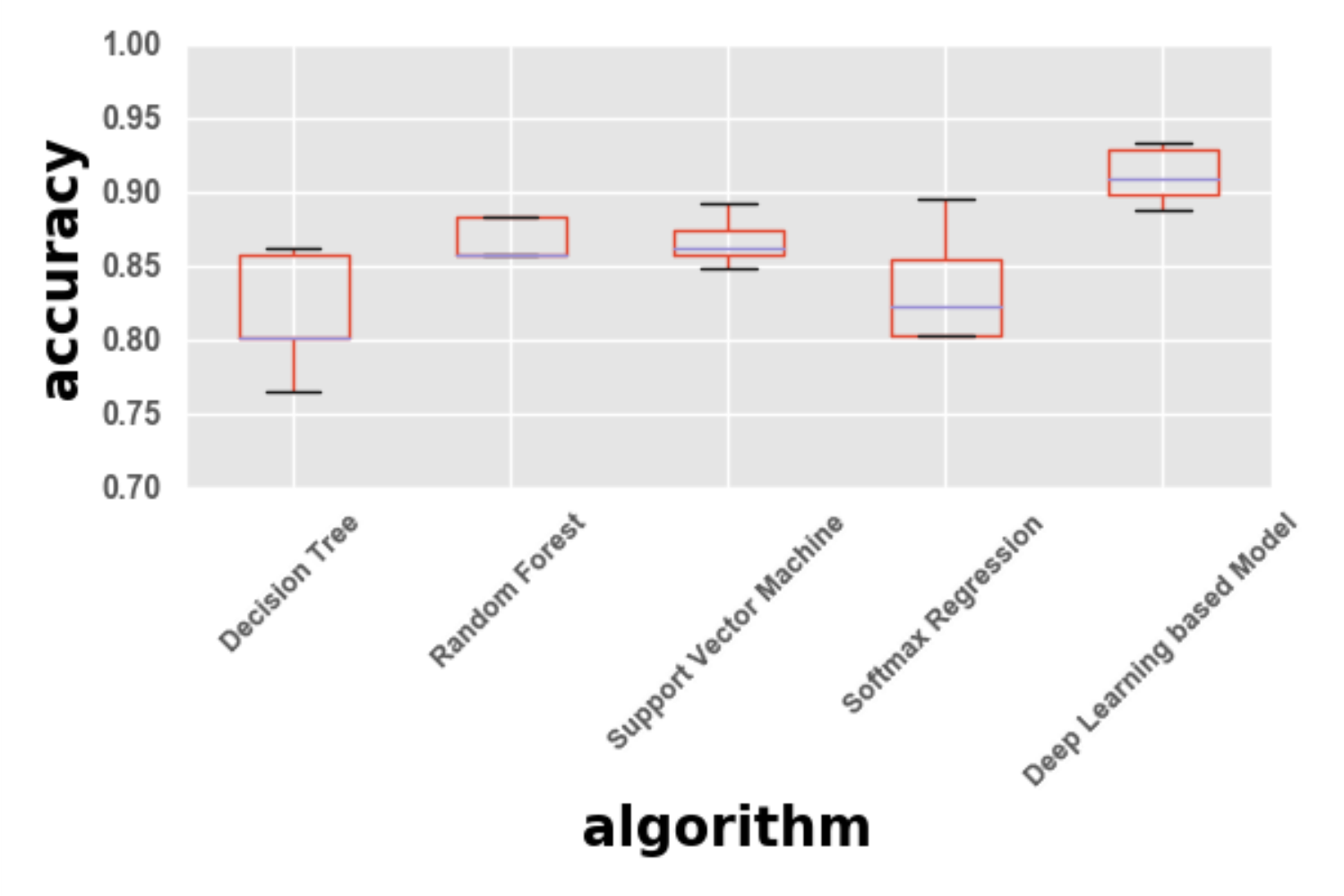}
  \caption{Accuracy distribution over 10 iterations for each algorithm.  In this boxplot, the wider box indicates more stable accuracy for 10 iterations.}
  \label{fig:algorithm-accuracy-boxplot}
\end{figure}

\begin{figure}[!ht]
  \centering
  \includegraphics[width=3.5in]{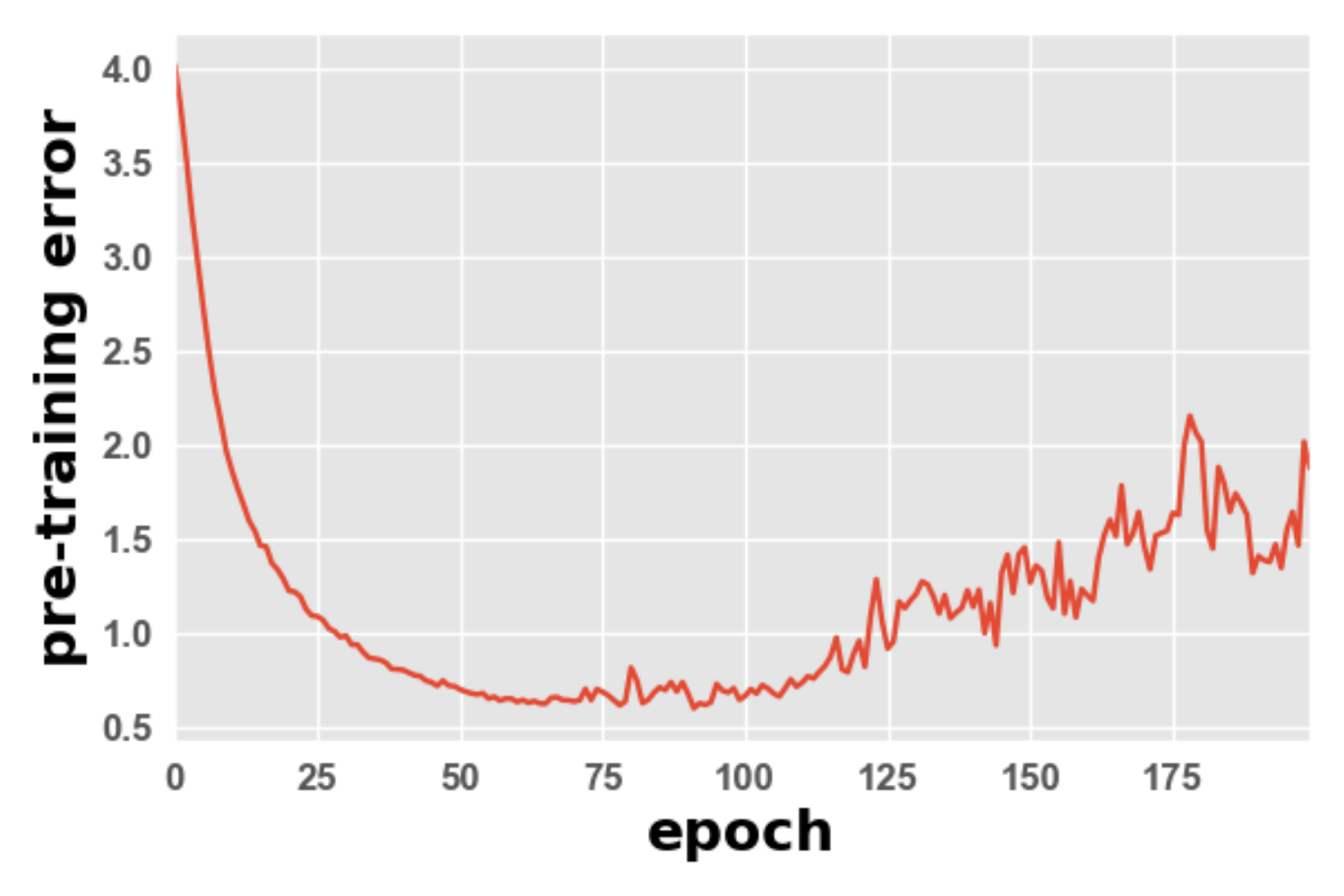}
  \caption{Pre-training errors on different epochs.  When epoch=50, the error reaches the minimum, and the effect of the pre-training is the best.}
  \label{fig:pre-train-error}
\end{figure}

\begin{figure}[!ht]
  \centering
  \includegraphics[width=3.5in]{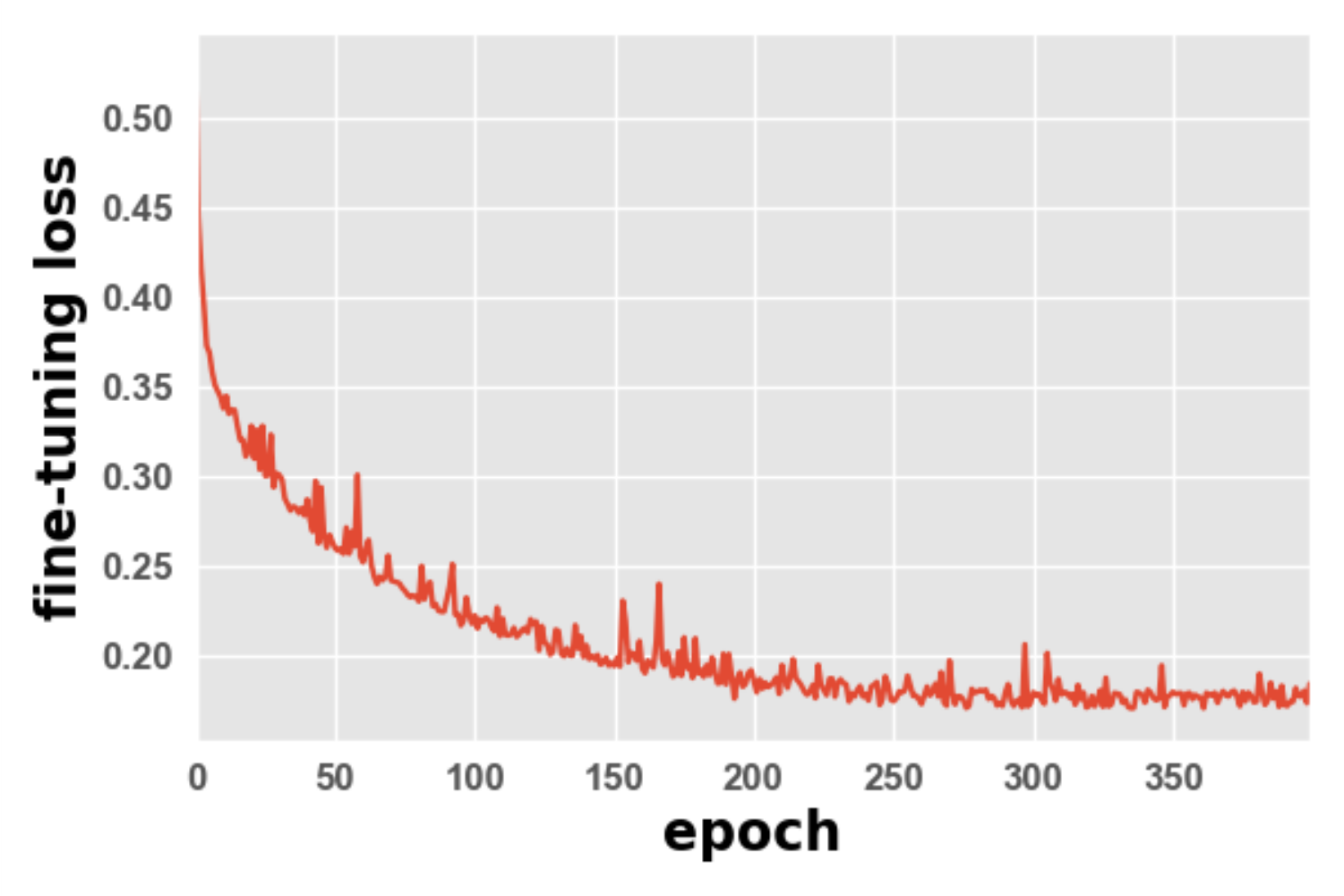}
  \caption{Fine-tuning losses on different epochs.  After epoch=350, the change in the loss value is very small.}
  \label{fig:fine-tuning-loss}
\end{figure}

In this subsection, we show three kinds of experimental results: (i) accuracy comparison of five algorithms on 10 datasets (Figure~\ref{fig:ratio-accuracy}), (ii) accuracy distribution over 10 iterations for each algorithm (Figure~\ref{fig:algorithm-accuracy-boxplot}), and (iii) pre-training errors on different training epochs (Figure~\ref{fig:pre-train-error}) and fine-tuning losses on different training epochs (Figure~\ref{fig:fine-tuning-loss}).

In these results, accuracy is the measurement of how well an algorithm can detect the malicious applications.  It is the proportion of all samples that are correctly classified, i.e., accuracy$:=\frac{\text{number~of~correctly~classified~samples}}{\text{total~number~of~samples}}$.  Accuracy distribution displays the distributed situation of the accuracy over 10 iterations.  For model training, in the pre-training phase, the mean-squared error measures the error from training RBMs with unlabeled samples, so it is the error on training with data, and in the fine-tuning phase, the categorical cross-entropy loss is used to measure the accuracy of classification.

From Figure~\ref{fig:ratio-accuracy}, it is clear that the proposed deep-learning-based model outperforms four other state-of-the-art algorithms in terms of the accuracy of detection.  It has 12.61\% higher accuracy compared with softmax regression, 5.76\% gain compared with decision tree, 3.20\% gain compared with support vector machine, and 2.61\% gain compared with random forest.  Moreover, by increasing size of the dataset, the model can realize better training to enhance the performance even further, whereas the four other algorithms do not show such dependency on the size of the dataset.

From Figure~\ref{fig:algorithm-accuracy-boxplot}, we can see that the accuracy distribution of the deep-learning-based model is almost always above the distributions of the four other algorithms.

Figure~\ref{fig:pre-train-error} clearly shows the model error after pre-training.  It initially decreases with an increase in the number of epochs, but after a certain time the error starts increasing, indicating that a certain epoch yields the minimum training error.  Upon increasing the training epoch, the training data are fitted into the model to let the model learn the features of the data.  The accuracy of the model is increased by such learning, but training longer than a certain duration starts increasing the error instead.

Figure~\ref{fig:fine-tuning-loss} shows that the loss in the fine-tuning stage decreases (although with some oscillations as the training duration increases).  In this training, there is the overfitting problem.

\subsection{Discussion}
In this subsection, we briefly discuss the generic insights into the challenges of deep-learning-based detection methods.

\textbf{Challenges:}  In this study, the most significant and critical challenge is handling streaming and fast-moving input data and to use these data to train the deep-learning-based model.  This challenge has two important aspects: (i) streaming and fast-moving data processing; and (ii) streaming-data-based model training.  In MEC, with the movement of mobile devices, streaming-data handling is an important ability to prevent and control attacks from malicious applications.  Moreover, it is time-consuming to train a deep-learning-based model, and it is important to train it online and to use it in real-time applications.

%======================================================================================================================================================================================================================================

\section{Conclusion}
\label{sec:conclusion}
In this paper, we have proposed a deep-learning-based model to detect malicious attacks in a mobile edge computing environment.  The automated feature learning characteristic of the deep-learning-based model was shown to be a significant advantage in terms of improving the accuracy of the malicious application detection rate compared with state-of-the-art machine-learning-based algorithms.  Moreover, we observed that the size of the training dataset plays an important role in further enhancing the accuracy of the deep-learning-based detection method.  On average, the accuracy of the proposed deep-learning-based model is 12.61\% higher compared with softmax regression, 5.76\% higher compared with decision tree, 3.20\% higher compared with support vector machine, and 2.61\% higher compared with random forest.

%======================================================================================================================================================================================================================================

%\ifCLASSOPTIONcompsoc
%\section*{Acknowledgments}

%======================================================================================================================================================================================================================================

\bibliographystyle{IEEEtran}
\bibliography{IEEEfull}

%======================================================================================================================================================================================================================================

\end{document}